\documentclass[usenatbib]{mnras}

\usepackage{epsfig}
\usepackage{amsmath, amssymb,bm}

\def \msun{\rm M_{\odot}}

\begin{document}
\title[Evading $M - \sigma$]{Supermassive Black Hole Demographics: Evading $M - \sigma$}

\author[Andrew King \& Rebecca Nealon ] 
{\parbox{5in}{Andrew King$^{1, 2, 3}$ \& Rebecca Nealon$^1$}
\vspace{0.1in} \\ $^1$ Department of Physics \& Astronomy, University
of Leicester, Leicester LE1 7RH UK\\ 
$^2$ Astronomical Institute Anton Pannekoek, University of Amsterdam, Science Park 904, NL-1098 XH Amsterdam, The Netherlands \\
$^{3}$ Leiden Observatory, Leiden University, Niels Bohrweg 2, NL-2333 CA Leiden, Netherlands}

\maketitle

\begin{abstract}
We consider black hole -- galaxy coevolution using simple analytic arguments. We focus
on the fact that several supermassive black holes are known with masses significantly larger than
suggested by the $M - \sigma$ relation, sometimes also with rather small stellar masses.
We show that these are likely to have descended from extremely compact `blue nugget' 
galaxies born at high redshift, whose very high velocity dispersions allowed the black holes to
reach unusually large masses. Subsequent interactions reduce the velocity dispersion, so the black holes
lie above the usual $M -\sigma$ relation and expel a large fraction of the bulge gas 
(as in WISE J104222.11+164115.3) that would otherwise make stars, before ending at low redshift 
as very massive holes in galaxies with relatively low stellar masses, such as NGC 4889 and NGC 1600. We further suggest the possible existence of two new types of galaxy: low--mass dwarfs whose
central black holes lie below the $M - \sigma$ relation at low redshift, and
galaxies consisting of very massive ($\ga 10^{11}\msun$) black holes with extremely small stellar masses. 
This second group would be very difficult to detect electromagnetically, but potentially offer targets of considerable
interest for LISA.
\end{abstract}


\footnotetext[1]{E-mail: ark@astro.le.ac.uk}

\section{Introduction}
\label{intro}
The realisation that observations imply 
scaling relations between supermassive black holes (SMBH) and their host galaxies
\citep[cf][]{Magorrian:1998,Haring:2004,Ferrarese:2000,Gebhardt:2000}
has stimulated significant efforts to identify the underlying physics (see \citealt{Kormendy:2013} and \citealt{King:2015} for reviews of observations and theory respectively). These have thrown up a range of ideas,
some involving the potential effects of merger averaging, or stellar feedback. But the fact that the
binding energy of the SMBH always far exceeds that of the host galaxy bulge strongly suggests 
black hole feedback as the basic cause.

In recent years a fairly coherent picture of SMBH 
growth and feedback in the local Universe has begun to emerge. It appears that low--redshift 
SMBH with a sufficient supply of surrounding 
gas grow their masses $M$ to the $M-\sigma$ value
\begin{equation}
M = M_{\sigma} \simeq \frac{f_g(1-f_g)\kappa}{\pi G^2}\sigma^4\simeq 
3\times 10^8f_g(1 - f_g) \sigma_{200}^4\msun
\label{msig}
\end{equation}
fixed by momentum--driven feedback from wide--angle winds driven off the SMBH accretion disc
in Eddington--limited phases SMBH \citep[cf][]{King:2003,King:2005}. Here
$\sigma = 200\sigma_{200}\,{\rm km\,s^{-1}}$ is the velocity dispersion of the
galaxy bulge,  $f_g$ is the gas fraction of baryonic matter in the bulge, $\kappa$ the electron 
scattering opacity, and $G$ the gravitational constant.
At this mass, SMBH feedback becomes energy--driven, and
drives away the remaining bulge gas, with typical velocity \citep{King:2005,Zubovas:2012}
\begin{equation}
v_{\rm out} \simeq 
1230\sigma_{200}^{2/3}l^{1/3}\,{\rm km\,s^{-1}}
\label{vout}
\end{equation}
where $l$ is the ratio of the driving SMBH accretion luminosity to the Eddington value.
This generally halts significant further SMBH growth, which can only resume if 
events such as mergers or dissipation rebuild
the bulge gas and restart evolution towards the Equation~\ref{msig} with larger $M$ and 
$\sigma$. Some growth beyond the Equation~\ref{msig} can occur if the SMBH has a low spin
and so is relatively inefficient in driving gas away (Zubovas \& King 2019, submitted), but this is 
relatively small.

In the simple picture described above, SMBHs can only grow significantly above $M_{\sigma}$ through mergers 
with other holes, or by sub--Eddington accretion exerting little feedback. Both these routes 
suggest that SMBH masses $M$ should be at most only slightly larger than 
$M_{\sigma}$, which is presumably why the $M - \sigma$ relation is observable at all.  
It is hard to find masses for SMBH below the Equation~\ref{msig}, because 
of the need to resolve their sphere--of--influence radii $R = 2GM/\sigma^2$ \citep[cf][]{Batcheldor:2010},
so it is possible that the $M - \sigma$ relation is really an upper limit to $M$ for given $\sigma$. Recent observational evidence \citep[e.g.][]{Yang:2019} out to redshifts $z = 3.0$ does suggest that the SMBH 
accretion rate and the star formation rate are effectively linearly related, which would maintain the scaling
relations specifically rather than as upper limits.

Despite the above discussion, there is clear evidence that $M$ is significantly larger than expected from the $M-\sigma$ relation, particularly at large $M$. The
most compelling examples are NGC 4889 \citep{McConnell:2011} and NGC 1600 \citep{McConnell:2011,Thomas:2016} in the centres of the Coma and Leo clusters respectively. 
These have SMBH masses $M = 2\times 10^{10}\msun$ and $1.7\times 
10^{10}\msun$. But using Equation~\ref{msig} even with the most favourable value $f _g = 1/2$, 
their observed velocity dispersions $\sigma = 350\,
{\rm  km\,s^{-1}}$ and  $293\, {\rm  km\,s^{-1}}$ give 
$M_{\sigma} = 7\times 10^9\msun, 3\times 10^9\msun$ respectively. For more likely values $f_g \sim 0.16$ the
predicted masses are $3.5\times 10^9\msun$ and $1.5\times 10^9\msun$, an order of magnitude
below the measured masses. \citet{Kormendy:2013} refer to such objects as `monsters', and note that
their discrepancies from the SMBH mass -- bulge mass relation $M \simeq 10^{-3}M_{\rm bulge}$
\citep{Haring:2004} are even larger than the offset from $M - \sigma$, in the sense that the bulge
mass is considerably reduced compared with this relation.
There are also several very large measured SMBH masses in the sample shown in 
\citet{Matsuoka:2018} Fig 4, and references cited there. These have no measured
$\sigma$, but the claimed
masses would require extremely large  values $\sigma \sim 500 - 1000\, {\rm km\,s^{-1}}$
to be compatible with Equation~\ref{msig}.  Evidently the coevolution of black holes and galaxies 
is more complex than the simple picture sketched above, and our aim here is to clarify this.

\section{SMBH Mass Evolution}

SMBH mass growth is controlled by three critical masses. These are 
$M_{\sigma}$ as defined above,  
the gas mass $M_g $ in the host galaxy bulge, and the
maximum SMBH mass $M_{\rm max}$ 
allowed if the SMBH is to have a luminous accretion disc and so exert feedback.
Equation~\ref{msig} gives $M_{\sigma}$, 
and we take the simple isothermal approximation
\begin{equation}
M_g = \frac{2f_g\sigma^2}{G}R
\label{mg}
\end{equation}
for $M_g$. The mass $M_{\rm max}$ is given by
requiring the black hole ISCO to be just equal to the disc self--gravity radius. At this
mass the disc breaks up into stars, severely inhibiting gas accretion on to the hole and the
resultant feedback on the host galaxy.   
This gives \citep{King:2016}
\begin{equation}
M_{\rm max} = 5\times 10^{10}\msun\alpha_{0.1}^{7/13}\eta_{0.1}^{4/13}
({\it L/L}_{\rm Edd})^{-4/13}{\it f}_5^{-27/26} .
\label{mmax}
\end{equation}  
Here $\alpha_{0.1}$ is the standard viscosity parameter in units of 0.1,
$\eta_{0.1}$ is the black hole accretion efficiency in units of 0.1, $L$ and $L_{\rm Edd}$ 
are respectively the accretion luminosity and Eddington
luminosity of the black hole, and
$f_5(a)$ is the spin dependence of the the innermost stable circular orbit 
(ISCO) radius in units of a typical value $5GM/c^2$, 
corresponding to a prograde spin $a \sim 0.6$. The full
range $-1< a < 1$ allows $M_{\rm max}$ values from  $2\times 10^{10}\msun$ up to
$3\times 10^{11}\msun$ \citep[][Fig 1]{King:2016}. SMBH mass growth beyond 
$M_{\rm max}$ is still possible -- indeed quite likely -- (see subsection 2.3 below), but
SMBH with masses $M > M_{\rm max}$ cannot exert feedback on their host galaxies.

The limit in Equation~\ref{mmax}
appears to be compatible with all extant observations. The SMBH in NGC 1600 is significantly below it for 
any spin rate, and this is probably 
true of NGC 4889 \citep[cf][]{King:2016}. A contender for the most extreme system is currently the very dust--obscured 
WISE J104222.11+164115.3, at redshift $z = 2.52$
\citep{Matsuoka:2018}. This galaxy has strongly blueshifted oxygen lines corresponding to an 
outflow velocity $\sim 1100\,{\rm km\, s^{-1}}$, close
to the prediction  in Equation~\ref{vout}. A luminous accretion disc is required to launch this, and
from Equation~\ref{mmax} the claimed SMBH mass $M = 10^{11}\msun$ requires it to have a spin $a 
\gtrsim 0.7$ and accrete prograde at the time of launch. We note that some caution is in order with this estimate, as the nature of the
blowout phase may affect the reliability of the usual virial mass estimators.

For simplicity in treating these effect in the analysis that follows we consider a fixed value 
\begin{equation}
M_{\rm max} = 3\times 10^{10}\msun,
\label{mmax2}
\end{equation}
corresponding to a spin parameter $a$  close to zero. This choice does not qualitatively affect our results,
as we shall see. 

For a given gas fraction $f_g$ we can write the ratios of these three critical 
masses in the simple forms
\begin{equation}
\frac{M_{\sigma}}{M_g} = 
\left(\frac{\sigma}{\sigma_0}\right)^2\frac{R_0}{R}
\label{sig/gas}
\end{equation}
\begin{equation}
\frac{M_g}{M_{\rm max}} = \left(\frac{\sigma}{\sigma_0}\right)^2\frac{R}{R_0}
\label{max/gas}
\end{equation}
\begin{equation}
\frac{M_{\sigma}}{M_{\rm max}} = \left(\frac{\sigma}{\sigma_0}\right)^4
\label{sig/max}
\end{equation}
where $\sigma_0$ and $R_0$ depend on $f_g$. We take 
$f_g = 0.5$ for `gas--rich' cases, and
$f_g = f_c \simeq 0.16$ 
for typical `cosmological' conditions. These choices give
$(\sigma_0, R_0) = (564\,{\rm km\, s^{-1}},\, 0.42\, {\rm kpc})$ 
$(\sigma_0, R_0) = (658\,{\rm km\, s^{-1}},\, 0.98\, {\rm kpc})$ for the 
gas--rich and cosmological cases respectively.

These values show that we can expect deviations from the usual low--redshift condition $M \lesssim M_{\sigma}$
only in galaxies born with very high velocity dispersions $\sigma \gtrsim 550\,{\rm km\, s^{-1}}$. But values
of $\sigma$ like this 
are required for galaxy bulges to have acquired large gas masses $\sim 10^{12}\msun$ (close to the upper limit
for rapid infall; \citealt{Rees:1977,White:1991}) by the redshifts $z \gtrsim 6$
where large SMBH masses are observed, as the gas cannot assemble faster than the dynamical rate 
\begin{equation}
\dot M_{\rm dyn}\sim 
\frac{f_g\sigma^3}{G}. 
\label{dyn}
\end{equation}
This infall only operates after cosmological outflow has slowed sufficiently (the turnround time) so typically for
a fraction $\sim 0.2$ of the lookback time $t_{\rm look}\simeq 10^9$\, yr at this redshift. 
Then the requirement 
\begin{equation}
\frac{f_g\sigma^3}{G}\times 0.2t_{\rm look} \gtrsim 10^{12}M_{12}\msun
\label{infall}
\end{equation}
for galaxies of mass $10^{12}M_{12}\msun$ gives an absolute lower limit, assuming continuous
accretion, of
\begin {equation}
\sigma \gtrsim 350 - 460 \times M_{12}^{1/3}\,{\rm km\,s^{-1}} 
\label{sigmacos}
\end{equation}
for $f_g = 0.5 -  0.16$. Simulations \citep[e.g.][]{vanderVlugt:2019} do indeed find large dispersions
for massive galaxies at this epoch. 


The three relations in Equations~\ref{sig/gas},~\ref{max/gas} and~\ref{sig/max} divide the $\sigma - R$ plane into 
the six regions $1 - 6$ shown in Figure \ref{demofig1}. We can now use it to see how SMBH grow in various cases. 

\subsection{SMBH growth for given $\sigma, R$}
\begin{figure*}
\centering
\includegraphics[width=0.8\textwidth]{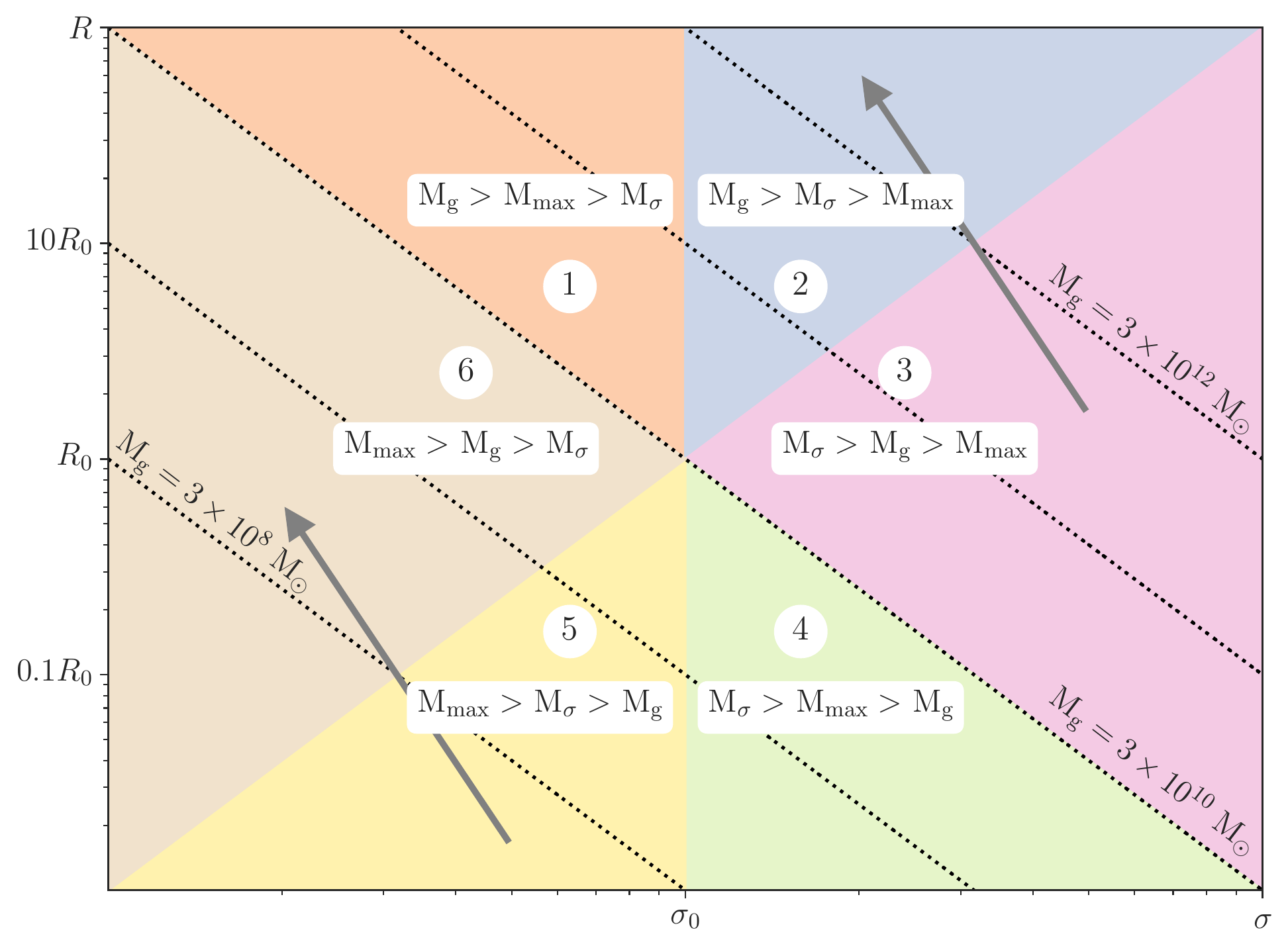}
\caption{The $(\sigma,\, R)$ plane for galaxy bulges and their central black holes. The 
orderings between the three critical masses $M_{\sigma}$, $M_g$ and $M_{\rm max}$ 
(cf Equations \ref{sig/gas}, \ref{max/gas}, \ref{sig/max}) divide this into regions $1 - 6$ as shown.  
The effect of minor mergers is to move a galaxy in the direction of the thick arrows.
At low redshift most galaxies are observed with SMBH masses $M \la M_{\sigma}$ 
and modest velocity dispersions $\sigma$
in the `normal' regions 1 and 6, depending on their total bulge gas masses $M_g$. Small
but very dense galaxies (`blue nuggets') are found at high redshift and necessarily with high $\sigma$ 
(cf Equation (\ref{sigmacos})) in regions 2 or 3. These
can grow large SMBH masses $M$ which are below $M_{\sigma}$ in those regions, but 
exceed the lower values of $M_{\sigma}$ prevailing 
in regions 1 if minor mergers or other processes such as dissipation move them there. 
This offers possible evolutionary 
routes to forming low--redshift galaxies whose SMBH have $M > M_{\sigma}$, such as 
NGC 1600 and NGC 4889, probably via systems like WISE J104222.11+164115.3, whose
vigorous feedback despite its very high SMBH mass suggests that 
$M_{\rm max} > M > M_{\sigma}$ here also. 
The figure is drawn for a single value of 
$M_{\rm max} = 3\times 10^{10}\msun$. Relaxing this to include the full range $2\times 
10^{10}\msun - 3\times 10^{11}\msun$ leaves the global behaviour of galaxies 
qualitatively unchanged, but blurs the boundaries between the regions $1 - 6$.}
 \label{demofig1}
\end{figure*}

Initially we consider how black holes grow in galaxy bulges in 
each region of Fig \ref{demofig1}, and consider the effects of evolution 
across the $(\sigma, R)$ plane later. 

In region 1 we have the familiar 
hierarchy $M_g > M_{\rm max} > M_{\sigma}$, so
SMBH growth in this region terminates at $M = M_{\sigma}$, as expected in the simple 
picture sketched in Section~\ref{intro}. SMBH feedback removes
the excess gas ($M_g > M_{\sigma}$) once $M$ reaches $M_{\sigma}$, 
and further growth requires an increase in the bulge velocity dispersion.
In region 6, $M_{\sigma}$ is again the smallest of the the three masses, so here too SMBH growth
stops at $M = M_{\sigma}$. Regions 1 and 6 differ only
in the overall mass of the galaxy, which is evidently larger in region 1 than 6. They contain the 
majority of galaxies observed at low redshift, almost all having $M \leq M_{\sigma}$. 

In regions 
4 and 5, the gas mass $M_g$ is smaller than the other two critical masses, so SMBH
growth stops before either $M_{\sigma}$ or $M_{\rm max}$ is reached. At low redshift
these galaxies would appear as low--mass dwarfs with black hole masses $M < M_{\sigma}$.
These masses must lie below the flattened $M - \sigma$ and $M - M_b$ (where $M_b$ 
is the bulge mass) relations found
by \citet{Martin-Navarro:2018} and \citet{Reines:2015}, but their possible 
presence agrees with the suggestion (from semi--analytic modelling)  by \citet{Pacucci:2018} 
of small central BH masses in dwarf galaxies.

The most 
interesting regions on Fig~\ref{demofig1} are 2 and 3, where $M_{\rm max}$ is the smallest of the
three critical masses, and so always smaller than $M_{\sigma}$. 
This means that feedback by itself cannot prevent SMBH growth, in principle all the way up to 
$M_{\rm max}$ and beyond (see below). What actually happens depends ultimately on what 
physical processes drive gas accretion on to SMBH, and how this relates to other galaxy properties, 
which is currently not fully understood.

If the black hole reaches mass $M_{\rm max}$, there is even less of a barrier to further SMBH growth
since accretion now produces neither radiation nor feedback, and there is no viscous accretion
disc to slow accretion either. We consider this in subsection 2.3.

\subsection{SMBH growth in evolving galaxies}

So far we have considered SMBH growth in galaxy bulges which evolve too little to move between the regions of 
Fig \ref{demofig1}. But in reality there are several ways that galaxy evolution can affect SMBH growth, such as
the processes of transformation and dissipation \citep{Bezanson:2012a}, and mergers with other
galaxies. In bulges undergoing transformation, star formation is quenched and $\sigma$ is likely to 
remain fixed. Since the long--term average SMBH accretion rate 
correlates with star formation rate, at least for bulge--dominated galaxies \citep[e.g.][]{Yang:2019} 
this presumably means that SMBH 
also stop growing. This is reasonable, since the SMBH mass $M$ and the bulge stellar mass $M_b$ are both 
proportional to
$\sigma^4$ (cf \citealt{Power:2011} and \citealt{King:2015}, Section 6). So these quiescent galaxies are
likely to remain on both the Faber--Jackson \citep{Faber:1976} and $M -\sigma$ relations, and therefore also on the SMBH --
bulge mass relation $M = \lambda M_b$, where $\lambda$ is constant at a value between 
$10^{-2}$ and $10^{-3}$ if average SMBH accretion and 
star formation track each other (\citealt{Yang:2019}: note that the treatment in \citealt{Power:2011} left the relation between these two quantities unspecified).

In contrast, dissipation causes gas to sink to the centre of the bulge, increasing the central gravitating mass and so 
$\sigma$, and decreasing the effective radius $R$. Similarly if a galaxy gains mass in minor mergers with other 
galaxies this reduces the velocity 
dispersion as $\sigma \sim M_g^{-1/2}$  (\citealt{Bezanson:2009}; the dispersion is likely to remain 
unchanged in major mergers -- cf \citealt{Marian:2019}). 

Using Equation~\ref{mg} we see this means that minor mergers move galaxies along tracks $\sigma 
\propto R^{-1/4}$, as shown in Fig \ref{demofig1} with the grey arrows. So either dissipation or mergers can reduce an initially large
$\sigma$ below  $\sigma_0$, particularly in the central regions near the SMBH,
moving galaxies from regions 2 and 3 into region 1. Here
the large SMBH mass grown in regions 2 and/or 3 may exceed the new, lower value of 
$M_{\sigma}$ specified by the central velocity dispersion. This will allow powerful SMBH 
feedback to restart, provided that $M$ has not yet reached 
$M_{\rm max}$. Full expulsion of the gas from the galaxy probably has to wait until $\sigma$ in the outer
regions drops below the critical value, leaving $M$ above the $M - \sigma$ value defined by the smaller central velocity dispersion. 

\begin{figure}
\centering
\includegraphics[width= 8 cm]{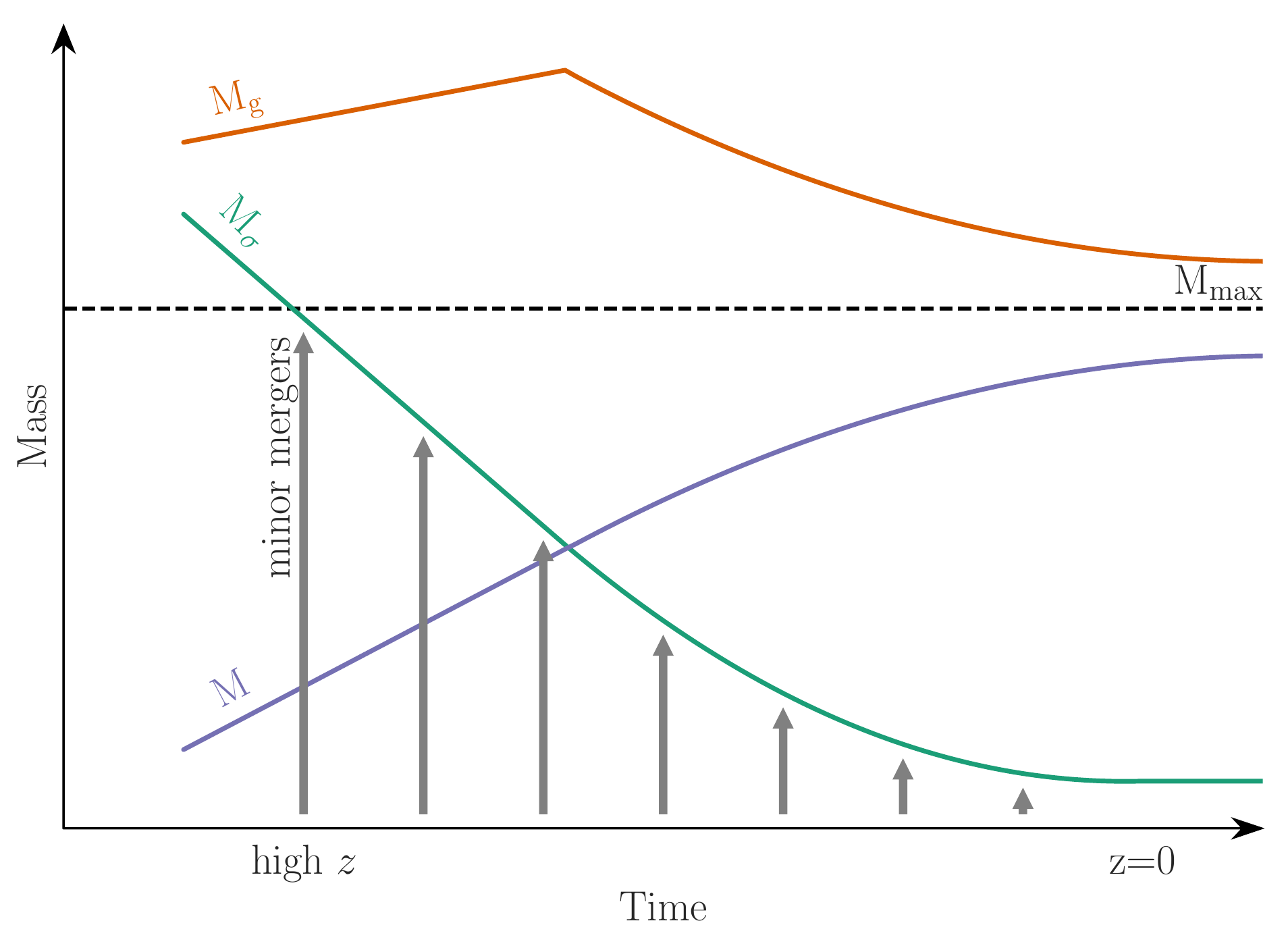}
\caption{Schematic view of the evolution producing galaxies such as NGC 4889 and NGC 1600. Minor
mergers gradually reduce the central velocity dispersion, reducing the critical mass $M_{\sigma}$ and 
leaving the SMBH mass $M$ 
above the $M - \sigma$ relation,
while feedback from SMBH accretion removes much of the gas mass $M_g$ and prevents the growth of a large stellar population.}
\label{demofig2}
\end{figure}

This evolution fits that of the system
WISE J104222.11+164115.3 \citep{Matsuoka:2018} discussed above. This appears to be
in the blowout phase, which would normally require its SMBH
mass $M \simeq 10^{11}\msun$ to be close to $M_{\sigma}$. But even
for a gas--rich galaxy the required value $\sigma \sim  1208\, {\rm km\, s^{-1}}$ is extremely high, so it is likely 
instead that $M$ is significantly greater than 
$M_{\sigma}$, as expected if it has evolved from
regions 2 or 3 into region 1. 
An evolution like this ultimately 
produces galaxies like NGC 4889 and NGC 1600. Because the bulge is largely gas, the strong 
SMBH feedback significantly depletes it before it can form stars, 
leaving the black hole 
marooned at a mass $M$ such that $M_{\rm max} > M >  M_{\sigma}$, together 
with a rather low stellar 
mass for the whole galaxy. This suggests a plausible 
evolutionary track from formation in regions 2 or 3, through galaxies with very high SMBH masses and currently in the blowout phase like WISE J104222.11+164115.3, ending as systems like NGC 4889 and NGC 1600. 
This is shown schematically in Fig \ref{demofig2}. 

 
\subsection{SMBH Growth Beyond $M_{\rm max}$?}

\begin{figure}
\centering
\includegraphics[width= 8 cm]{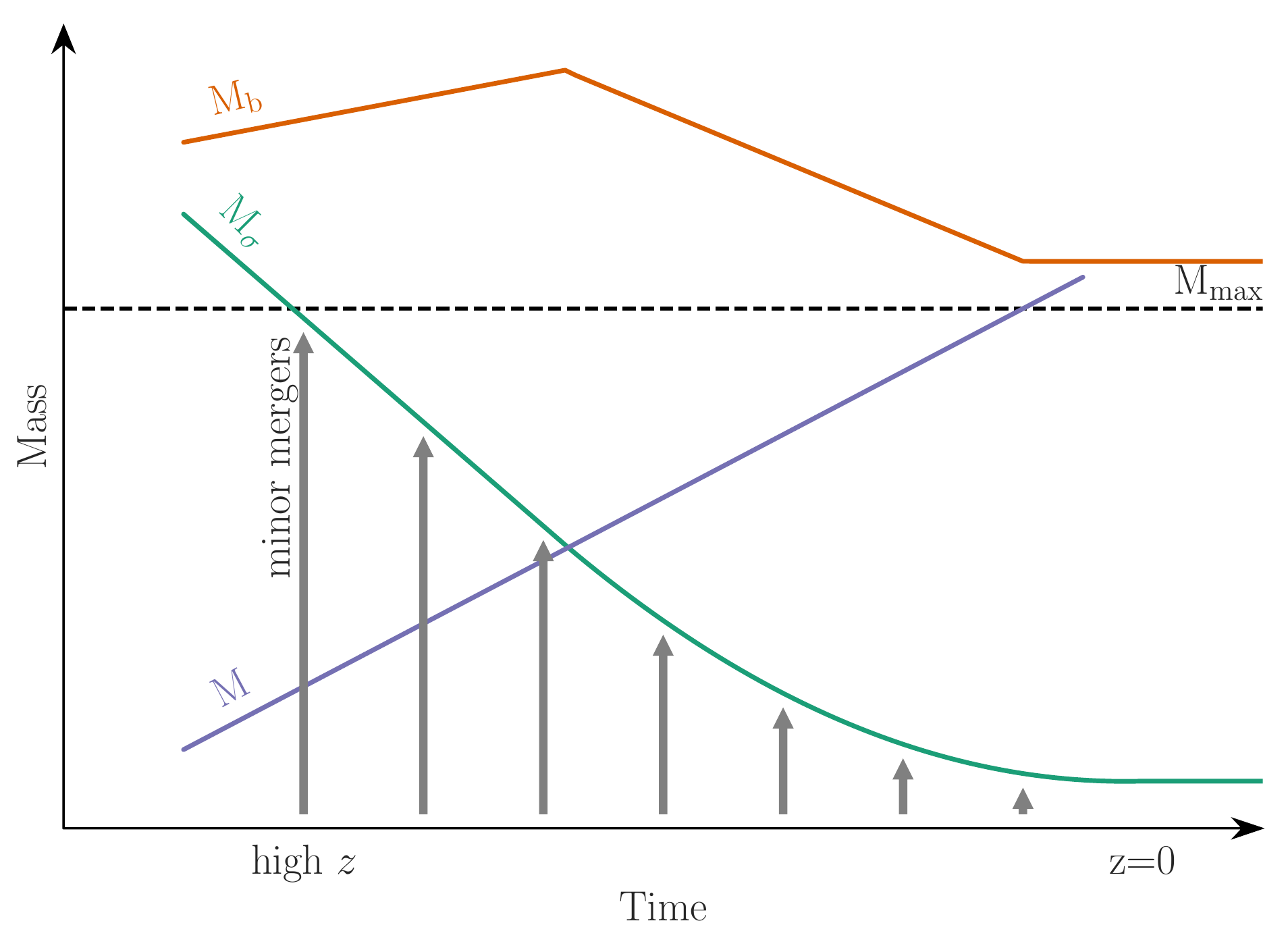}
\caption{Schematic view of the evolution producing galaxies with very massive SMBH and a minimal stellar
component. This is similar to that of Fig. (\ref{demofig2}), but here the SMBH evolves beyond $M_{\rm max}$. 
These objects would be almost undetectable by electromagnetic observations, but potentially 
observable by LISA.}
\label{demofig3}
\end{figure}

We emphasized above that $M_{\rm max}$ is only a limit for the mass of an SMBH undergoing 
luminous accretion from a disc. 
Given a continuing mass supply,  a SMBH with $M \geq M_{\rm max}$ can quietly swallow 
whole some of the stars forming near the ISCO, as it is well above the limit for tidally disrupting them.
It can grow its mass even more easily than before, as it is now 
freed of constraints such as the Eddington limit or the $M - \sigma$ mass, which are only 
relevant if accretion produces feedback. Fig 1 of King (2016) shows that several SMBH are very 
close to this regime. In a gas--rich case this process could leave the galaxies with very 
low stellar masses, and an extremely massive black hole (cf Fig. \ref{demofig3}).
They would be effectively undetectable by 
conventional (electromagnetic) means, but clearly targets of great interest for LISA, most spectacularly through majors mergers, but also through extreme mass ratio inspirals (EMRIs).

\section{Conclusions}

We have shown that galaxies formed by rapid gas infall at redshifts $z \ga 6$ may grow central 
black holes with very large masses. As their velocity dispersions decrease over cosmic time, feedback
from the SMBH can remove a significant fraction of the gas otherwise available for star formation.
At low redshift these galaxies have SMBH well above the usual
$M - \sigma$ and the SMBH -- bulge mass relations.
These properties agree with those of NGC 4889 and NGC 1600, and probably other galaxies too.
We point out the possible existence of low--mass dwarf galaxies whose central black holes lie
significantly below the $M - \sigma$ relation, and galaxies with very massive black holes but extremely 
small stellar masses. Electromagnetic identification of members of this second group would be very difficult, 
but they are potentially very interesting for LISA.

\section*{Acknowledgments}
The authors thank Kastytis Zubovas for valuable comments and the referee for a very helpful report. This project has received funding from the European Research Council (ERC) under the European Union's Horizon 2020 research and innovation programme (grant agreement No 681601).

\bibliographystyle{mnras}
\bibliography{bibfile} 

\end{document}